\documentclass[twocolumn,showpacs,floatfix,aps]{revtex4}
\usepackage{graphicx}
\usepackage{dcolumn}
\usepackage{bm}
\def\tr{t_{\rm r}}

\begin{document}
\title{Activated escape of periodically modulated systems}

\author{M.I. Dykman and D. Ryvkine }
\affiliation{Department of Physics and Astronomy, Michigan State University}
\date{\today}
\begin{abstract}
The rate of noise-induced escape from a metastable state of a
periodically modulated overdamped system is found for an arbitrary
modulation amplitude $A$. The instantaneous escape rate displays peaks that
vary with the modulation from Gaussian to strongly asymmetric. The
prefactor $\nu$ in the period-averaged escape rate depends on $A$
nonmonotonically.  Near the bifurcation amplitude $A_c$
it scales as $\nu\propto (A_c-A)^{\zeta}$. We identify three scaling
regimes, with $\zeta = 1/4, -1$, and $1/2$.
\end{abstract}


\maketitle

Thermally activated escape from a metastable state is often investigated in
systems driven by time-dependent fields. Recent examples are activated
transitions in modulated nanomagnets \cite{Wernsdorfer97,Koch00,Ralph02} and
Josephson junctions \cite{Han03,Wallraff_nature03,Siddiqi04}. Modulation
changes the activation barrier. This enables efficient control of the escape
rate and accurate measurement of the system parameters
\cite{Devoret87,Cleland04}.

Most frequently used types of modulation are slow ramping of a control
parameter, when the system remains quasistationary, and periodic
modulation. In the latter case the system is away from thermal
equilibrium, which  complicates the theoretical
formulation of the escape problem \cite{Soskin03}.

In the present paper we extend to periodically modulated systems the analysis
of the escape rate done by Kramers for systems in thermal equilibrium
\cite{Kramers}. Our approach gives the full time-dependent escape rate $W(t)$
as well as the period-averaged rate $\overline W=\nu\exp(-R/D)$, where $R$ is
the activation energy of escape and $D$ is the noise intensity, $D=k_BT$ for
thermal noise.


For comparatively small modulation amplitude $A$ escape of an
overdamped Brownian particle was studied in Ref.~\onlinecite{SDG}. The
range of intermediate $A$ and intermediate modulation frequencies
$\omega_F$ was analyzed in Refs.~\onlinecite{Hanggi-00,M&S-01}. Here
we find $W(t)$ for an arbitrary $A$ and an arbitrary interrelation
between $\omega_F$ and the relaxation time of the system $t_{\rm
r}$. We show that the prefactor $\nu$ depends on $A$ strongly and
nonmonotonically. It displays scaling behavior near the bifurcational
modulation amplitude $A_c$ for which the metastable state disappears.

In the spirit of Kramers' approach, we relate the instantaneous escape
rate $W(t)$ to the current {\it well behind} the boundary $q_b(t)$ of
the basin of attraction to the initially occupied metastable state
($q$ is the system coordinate). This is the current usually studied in
experiments. Because of the oscillations of $q_b(t)$, it has a
different functional form from the current at $q_b(t)$ calculated in
Refs.~\onlinecite{Hanggi-00,M&S-01}.

We find $W(t)$ by matching the probability distribution $\rho(q,t)$
near $q_b(t)$ and inside the basin of attraction.  This can be done in
a broad parameter range without a complete calculation of $\rho(q,t)$
near $q_b(t)$, using singular features of the dynamics of large
fluctuations.

For a periodically modulated overdamped Brownian particle, the
distribution $\rho(q,t)$ is given by the Fokker-Planck equation (FPE)
\begin{equation}
\label{FPE}
\partial_t \rho = -\partial_q\left[
K(q,t)\rho\right] + D\partial^2_q\rho.
\end{equation}
Here, $K(q,t)$ is the periodic force driving the particle,
$K(q,t)=K(q,t+\tau_F)\equiv-\partial_qU(q,t)$, where $\tau_F=2\pi/\omega_F$ is
the modulation period and $U(q,t)$ is the metastable potential.  The equation
of motion of the particle in the absence of noise is $\dot q = K(q,t)$.
%
The metastable state $q_a(t)$, from the vicinity of which the system
escapes due to noise, and the basin boundary $q_b(t)$ are the stable
and unstable periodic solutions of this equation, respectively.

We will assume that the noise intensity $D$ is small. Then in a broad
time range $\tr \ll t \ll 1/\overline W$ the distribution $\rho(q,t)$
is nearly periodic in the basin of attraction
to $q_a(t)$. The current away
from this basin, and thus the escape rate $W(t)$, are
also periodic.

The distribution $\rho$ is maximal at
$q_a(t)$ and falls off exponentially away from it. In the presence of
periodic driving it acquires singular features as $D\to 0$
\cite{Graham-84}, some of which have counterparts in wave fields
\cite{Berry-76}, with $D$ playing the role of the wavelength. The
singularities accumulate near $q_b(t)$. In order to find $W(t)$ one
has to understand how they are smeared by diffusion.

In the absence of noise the motion of the system close to the periodic
states $q_{i}(t) \;(i=a,b)$ is described by the equation $\dot q = K$
with $K$ linearized in $q-q_i(t)$. The
evolution of $q(t)-q_i(t)$ is given by the factors
\begin{equation}
\label{lambda}
\kappa_i(t,t')=\exp
\left[\int\nolimits_{t'}^td\tau\,\mu_i(\tau)\right] \quad (i=a,b),
\end{equation}
where $\mu_i(t) = \mu_i(t+\tau_F) \equiv [\partial_q
K(q,t)]_{q_i(t)}$.
Over the period $\tau_F$ the distance $q(t)-q_i(t)$ decreases (for $i=a$)
or increases (for $i=b$) by the Floquet multiplier
$M_i=\kappa_i(t+\tau_F,t) \equiv \exp(\bar\mu_i\tau_F)$, where
$\bar\mu_i$ is the period-average value of $\mu_i(t)$, with $\bar\mu_a
< 0, \bar\mu_b > 0$.

For weak noise the expansions of $K$ can be used to find $\rho(q,t)$
near $q_{a,b}(t)$. Near the metastable state $q_a$, the distribution
is Gaussian \cite{Ludwig}, $\rho(q,t)\propto
\exp\{-[q-q_a(t)]^2/2D\sigma^2_a(t)\}$. The reduced time-periodic
variance is given by the equation
\begin{equation}
\label{sigma_a}
\sigma^2_i(t)=2\left|M_i^{-2}-1\right|^{-1}\int\nolimits_0^{\tau_F}dt_1 \kappa_i^{-2}(t+t_1,t)
\end{equation}
with $i=a$ (in the absence of modulation $\sigma^2_a=1/|\mu_a|$).

The general form of the periodic distribution near the unstable state
$q_b(t)$ (the boundary-layer distribution) can be found from
Eq.~(\ref{FPE}) using the Laplace
transform, similar to the weak-driving limit \cite{SDG}. With $K$
linear in $q-q_b$, the equation for the Laplace transform of $\rho(q,t)$ is
of the first order, giving
\begin{eqnarray}
\label{near_q_b}
\rho(q,t) &=& \int\nolimits_0^{\infty}dp\,
e^{-pQ/D}\, \tilde\rho(p,t),\; Q=q-q_b(t), \nonumber\\
\tilde\rho(p,t) &=& {\cal E}D^{-1/2}\, \exp\left\{-\left[s(\phi)+
p^2\sigma^2_b(t)/2\right]/D\right\}.
\end{eqnarray}
In Eq.~(\ref{near_q_b}), ${\cal E}$ is a constant, $s(\phi)$ is an
arbitrary zero-mean periodic function, $s(\phi + 2\pi) = s(\phi)$, and
$\phi \equiv \phi(p,t)$,
\begin{equation}
\label{phase}
\phi(p,t) = \Omega_F \ln[p\,\kappa_b(t,t^{\prime})/\bar\mu_bl_D].
\end{equation}
Here, $\Omega_F=\omega_F/\bar\mu_b \equiv 2\pi/\ln M_b$ is the reduced
field frequency, $l_D=(2D/\bar\mu_b)^{1/2}$ is the typical diffusion
length, and $t^{\prime}$ determines the initial value of $\phi$; from
Eq.~(\ref{phase}), $\phi(p,t+\tau_F)=\phi(p,t)+2\pi$. In
Eq.~(\ref{near_q_b}) we assumed that the basin of attraction to $q_a$ lies for
$q<q_b(t)$, and $|Q|\ll \Delta q\equiv \min_t[q_b(t)-q_a(t)]$.


The form (\ref{near_q_b}) is advantageous as it immediately gives the
current $j(q,t)$ from the occupied region $(-\infty,q]$. Well behind
the basin boundary, where $Q= q-q_b(t)\gg l_D$, diffusion can be
disregarded, the current becomes convective and gives the
instantaneous escape rate. With $\dot q\approx \dot q_b+\mu_bQ$, we
have $j(q,t)\approx \mu_b(t)\rho(q,t)Q$, at a given $Q$. Disregarding
the term $\propto p^2/D$ in $\tilde \rho$ for $Q\gg l_D$, we obtain
from Eq.~(\ref{near_q_b})
\begin{equation}
\label{W_general} j(q,t)=\mu_b(t){\cal E}D^{1/2}\int\nolimits_0^{\infty}dx\,
e^{-x}\,\exp[-s(\phi_d)/D].
\end{equation}
Here, $\phi_d = \Omega_F \ln[x\,\kappa_b(t_d,t^{\prime})]$, and
$t_d\equiv t_d(Q,t)$ is given by the equation $\kappa_b(t_d,t) =
l_D/2Q$. In the whole harmonic range $j$ depends on the observation
point $Q$ only in terms of the delay time $t_d$, which shows how long
it took the system to roll down to the point $Q$, $\partial
t_d/\partial Q =- 1/\mu_b(t_d)Q$. We note that $\mu_b(t)$ can be
negative for a part of the period, leading to reversals of the
instantaneous current.

The escape rate $\overline{W}$ is given by the
period-averaged $j(q,t)$ and is independent of $q$. From
Eq.~(\ref{W_general})
\begin{equation}
\label{W_bar} \overline W=\frac{\bar\mu_b}{2\pi}{\cal
E}D^{1/2}\int\nolimits_0^{2\pi}d\phi\, \exp[-s(\phi)/D].
\end{equation}

Eqs.~(\ref{W_general}) and (\ref{W_bar}) provide a complete solution
of the Kramers problem of escape of a modulated system and reduce it
to finding the function $s$. They are similar in form to the
expressions for the instantaneous and average escape rates for
comparatively weak modulation, $|s|\ll R$, where $s$ was obtained
explicitly \cite{SDG}.

Unless the modulation is very weak or has a high frequency, for small
noise intensity $\max\, s\sim |\min\, s| \gg D$. In this case the
major contribution to the integrals in Eqs.~(\ref{W_general}),
(\ref{W_bar}) comes from the range where $s$ is close to its minimum
$s_{\rm m}$ reached for some $\phi=\phi_{\rm m}$. Then the escape rate
$j(q,t)$ sharply peaks as function of time once per period when
$\phi_d(t)=\phi_{\rm m}$. This means that escape events are {\it
strongly synchronized}. As we show, both $j(q,t)$ and $\overline W$
are determined not by the global shape of $s(\phi)$, but only by the
curvature of $s(\phi)$ near $\phi_{\rm m}$.

To find $j(q,t)$ we match Eq.~(\ref{near_q_b}) to the distribution
$\rho(q,t)$ close to $q_b(t)$ but well inside the attraction basin,
$-Q\gg l_D$. For small $D$ this distribution can be found, for
example, by solving the FPE (\ref{FPE}) in the eikonal approximation,
$\rho(q,t)= \exp[-S(q,t)/D]$. To zeroth order in $D$, the equation for
$S=S_0$ has the form of the Hamilton-Jacobi equation $\partial_t S_0 =
- H$ for an auxiliary nondissipative system with Hamiltonian
\cite{Freidlin-book}
\begin{equation}
\label{Hamiltonian}
H(q,p;t)= p^2 + pK(q,t), \; p = \partial_q S_0.
\end{equation}
The Hamiltonian trajectories of interest for escape $q(t),p(t)$ start
in the vicinity of the metastable state. The initial conditions follow
from the Gaussian form of $\rho(q,t)$ near $q_a(t)$, with $S_0
=[q-q_a(t)]^2/2\sigma^2_a(t)$.

To logarithmic accuracy, the escape rate is determined by the
probability to reach the basin boundary $q_b(t)$, i.e., by the action
$S_0\bigl(q_b(t),t\bigr)$ \cite{Soskin03}.  The Hamiltonian trajectory
$q_{\rm opt}(t), p_{\rm opt}(t)$, which minimizes
$S_0\bigl(q_b(t),t\bigr)$, approaches $q_b(t)$ asymptotically as $t\to
\infty$. This is a heteroclinic trajectory of the auxiliary system, it
is periodically repeated in time with period $\tau_F$; $q_{\rm
opt}(t)$ is the most probable escape path (MPEP) of the original
system.

Close to $q_b(t)$, the Hamiltonian equations for $q(t),p(t)$ can be
linearized and solved. On the MPEP
%
\begin{eqnarray}
\label{action_near_b}
&&p_{\rm opt}(t)=-Q_{\rm
opt}(t)/\sigma_b^2(t)=\kappa_b^{-1}(t,t^{\prime})p_{\rm opt}(t^{\prime}),\\
&&S_0\bigl(q_{\rm opt}(t),t\bigr) = R - Q_{\rm opt}^2(t)/2\sigma_b^{2}(t),\nonumber
%
\end{eqnarray}
where $Q_{\rm opt}(t)=q_{\rm opt}(t)-q_b(t)$. The quantity $R =
S_0\bigl(q_{\rm opt}(t),t\bigr)_{t\to \infty}$ is the activation
energy of escape.

The surface $S_0(q,t)$ is flat for small $Q-Q_{\rm opt}$ due to
nonintegrability of the dynamics with Hamiltonian (\ref{Hamiltonian})
\cite{Graham-84}. It touches the surface
$S_b(q,t)=R-Q^2/2\sigma_b^2(t)$ on the MPEP, $Q=Q_{\rm opt}(t)$. Away
from the MPEP $S_0(q,t)>S_b(q,t)$, and therefore the function
$\rho_b(q,t)=\rho(q,t)\exp[S_b(q,t)/D]$ is maximal on the MPEP.

We match on the MPEP $\rho_b$ found in the eikonal approximation to
the maximum of $\rho_b$ found from Eq.~(\ref{near_q_b}) near the basin
boundary. For $|s_{\rm m}| \gg D$ and $ -Q\gg l_D$, the integral over
$p$ in Eq.~(\ref{near_q_b}) can be evaluated by the steepest descent
method. The integrand is maximal if $p=-Q/\sigma_b^2(t)$ and $s$ is
minimal for this $p$, i.e., $\phi(p,t)=\phi_{\rm m}$ and $s=s_{\rm
m}$. These conditions can be met on the whole MPEP at once,
because $\phi(p_{\rm opt}(t),t)=$const. Then from
Eq.~(\ref{near_q_b})
\begin{eqnarray}
\label{steep_descent}
&&\rho(q,t)={\cal E}_b(t)\exp[-S_b(q,t)/D], \nonumber\\
&&{\cal E}_b(t)= \tilde{\cal E}D^{-1/2} \left[\sigma_b^2(t) + \Omega_F^2 \,
s''_{\rm m}\, p^{-2}_{\rm opt}(t)\right]^{-1/2},
\end{eqnarray}
where $\tilde{\cal E}={\cal E}(2\pi D)^{1/2}\exp[(R-s_{\rm m})/D]$,
and $s''_{\rm m}\equiv [d^2s/d\phi^2]_{\phi_{\rm m}}$. From
Eqs.~(\ref{action_near_b}), (\ref{steep_descent}), not only the
exponents, but also their slopes coincide along the MPEP for the
boundary-layer and eikonal-approximation distributions.

The function ${\cal E}_b(t)$ should match on the MPEP the prefactor of the
eikonal-approximation distribution $\rho=\exp(-S/D)$, which is given by the
term $S_1\propto D$ in $S$.  On the MPEP, $z=\exp(2S_1/D)$ obeys the equation
\cite{Hanggi-00,Ryter}
\begin{equation}
\label{G_equation} d^2z/dt^2 -2d(z\,\partial_qK)/dt + 2zp\partial^2_qK = 0,
\end{equation}
where $q=q_{\rm opt}(t),\, p=p_{\rm opt}(t)$. The initial condition to
this equation follows from $\rho(q,t)=z^{-1/2}\exp(-S_0/D)$ being
Gaussian near $q_a(t)$, which gives $z(t)\to 2\pi D\sigma_a^2(t)$ for
$t\to -\infty$.  Close to $q_b(t)$, from Eq.~(\ref{G_equation}) $z(t)
= D[\mathcal{Z}_1\sigma_b^2(t)+\mathcal{Z}_2p^{-2}_{\rm opt}(t)]$, where $\mathcal{Z}_{1,2}$ are
constants \cite{Hanggi-00}; the term $\propto\mathcal{Z}_1$ was disregarded in
the analysis \cite{Hanggi-00}.
Remarkably, $z^{-1/2}(t)$ is of the same functional form near $q_b(t)$
as ${\cal E}_b(t)$. Thus the prefactors in $\rho(q,t)$ as given by the
eikonal and the boundary-layer approximations also match each other.

Explicit expressions for the escape rate in the regime of strong
synchronization can be obtained for comparatively weak or slow
modulation, where $ s''_{\rm m} \sim |s_{\rm m}|\gg D$ but
\begin{equation}
\label{moderate}
\Omega_F^2s''_{\rm m} \ll R.
\end{equation}
The results for $D\ll |s_{\rm m}| \ll R$ should coincide with the
results of Ref.~\onlinecite{SDG}, which were obtained in a different
way. We have verified this by finding $s''_{\rm m}$ from
Eq.~(\ref{G_equation}) by perturbation theory in the modulation
amplitude $A$.

Condition (\ref{moderate}) can be met for large $A$, where $s''_{\rm
m}\sim R$, provided the modulation frequency is small, $\omega_Ft_{\rm r}\sim
\Omega_F\ll 1$ (adiabatic modulation). Here, the MPEP is given by the equation
$\dot q_{\rm opt}=-K\bigl(q_{\rm opt},t_{\rm m}\bigr)$, with $t_{\rm m}$ found
from the condition of the minimum of the adiabatic barrier height $\Delta
U(t)=U\bigl(q_b(t),t\bigr)-U\bigl(q_a(t),t\bigr)$. The activation energy
$R=\Delta U_{\rm m} \equiv \Delta U(t_{\rm m})$.

The value of $s''_{\rm m}$ can be obtained from $z(t)$ or by matching the
adiabatic intrawell distribution $\propto\exp[-U(q,t)/D]$ and the boundary
layer distribution (\ref{near_q_b}) in the region $|Q|\gg l_D$ and
$\Omega_F^2s''_{\rm m} \ll \mu_b(t_{\rm m})Q^2$  for $|t-t_m|\ll \tau_F$. Both
approaches give $\Omega_F^2\mu_b^2s''_{\rm m}=\Delta\ddot U_{\rm m}$, where
$\mu_{b}$ and $\Delta\ddot U_{\rm m}\equiv
\partial^2_t\Delta U$ are calculated for $t=t_{\rm m}$.

The form of $j(q,t)$ depends on the parameter $\Omega_F^2s''_{\rm
m}/D$. When it is small, the term $\propto p^{-2}_{\rm opt}$ in ${\cal
E}_b(t)$ [Eq.~(\ref{steep_descent})] and $z(t)$ is also small away from
the diffusion region around $q_b$. Then $z = 2\pi D\sigma_a^2(t_{\rm
m})$. The pulses of $j(q,t)$ are Gaussian,
\begin{equation}
\label{Gaussian_peaks}
j(q,t)=\frac{|\mu_a\mu_b|^{1/2}}{2\pi}e^{-R/D}\sum\nolimits_k
e^{-(t-t_k)^2\Delta\ddot{U}_{\rm m}/2D}
\end{equation}
[$\mu_{a,b} \equiv \mu_{a,b}(t_{\rm m})$]. They are centered at
$t_k=t_{\rm m}+k\tau_F$, with $k=0,\pm 1,\ldots$ [we disregard the
delay $\sim \mu_b^{-1}\ln (Q/l_D)$ in
$t_k$]. Eq.~(\ref{Gaussian_peaks}) corresponds to the fully adiabatic
picture, where the escape rate is given by the instantaneous barrier
height $\Delta U(t)$.

The current has a different form for $\Omega_F^2s''_{\rm m}/D \gg
1$. Because $ p^{-2}_{\rm opt}(t)\propto\kappa_b^2(t,t')$
exponentially increases in time near $q_b$, the term $\propto
p^{-2}_{\rm opt}$ in ${\cal E}_b$ and $z$ becomes dominating before
the MPEP reaches the diffusion region $|Q|\sim l_D$. Then
Eqs.~(\ref{W_general}), (\ref{steep_descent}) give
\begin{eqnarray}
\label{instant_rate} &&j(q,t) = \frac{\mu_b(t)\tilde{\cal
E}D^{1/2}}{\Omega_F\sqrt{s''_{\rm m}}}e^{-R/D}
\sum\nolimits_{k=-\infty}^{\infty} x_ke^{-x_k},\\
&& x_k=x_0\exp(2\pi k/\Omega_F),\qquad x_0 =p_{\rm opt}(t)Q/D. \nonumber
\end{eqnarray}
Note that here $p_{\rm opt}(t)$ can be smaller than $l_D/\sigma_b^2(t)$.

Eq.~(\ref{instant_rate}) describes the escape rate in the whole region
$\Omega_F^2s''_{\rm m} \gg D$; it does not require the adiabatic
approximation. Its form is totally different from that of the
diffusion current $-D\partial_Q\rho$  on the basin boundary
$Q=0$ as given by Eqs.~(\ref{near_q_b}), (\ref{steep_descent}).  The ratio
$\tilde{\cal E}/\sqrt{s''_{\rm m}}=\Omega_F\mathcal{Z}_2^{-1/2}$ can be obtained
by solving Eq.~(\ref{G_equation}).

For $\Omega_F\ll 1$ the current ({\ref{instant_rate}) is a series of
distinct strongly asymmetric peaks, with $x_k\approx
\exp[-(t-k\tau_F-t_{\rm m})\mu_b(t_{\rm m})]$ near the maximum. The
transition between the pulse shapes (\ref{Gaussian_peaks}) and
(\ref{instant_rate}) occurs for $\Omega_F^2s''_{\rm m}/D \sim 1$. It
is described by Eq.~(\ref{W_general}) with ${\cal E}=(2\pi)^{-1}
D^{-1/2}|\mu_a/\mu_b|^{1/2}\exp[-(R-s_{\rm m})/D]$. We note that, for
$\Omega_F\ll 1$, the shape of current pulses in the whole range
(\ref{moderate}) is the same as for weak modulation
\cite{Smelyanskiy_JPC99}, but the parameters depend on $A,\omega_F$
differently.

With increasing $\Omega_F$ the peaks of $j$ ({\ref{instant_rate}) are smeared
out and the escape synchronization is weakened. For $\Omega_F\gg 1$ it
disappears ($s''_{\rm m}$ rapidly decreases with $\omega_F$ for large
$\Omega_F$).

The escape current (\ref{instant_rate}) is completely different from
the current on the basin boundary~\cite{Hanggi-00}. The regime
$\Omega_{F}^{2}s''_{\rm m}/D \ll 1$, where the current has the form
(\ref{Gaussian_peaks}), cannot be studied in the approximation
\cite{Hanggi-00} at all.

In the range $s''_{\rm m}\sim |s_{\rm m}|\gg D$, the period-averaged
escape rate (\ref{W_bar}) is
\begin{equation}
\label{period_averaged}
\overline{W}=\nu\exp(-R/D),\qquad
\nu=\bar\mu_b\tilde{\cal E}D^{1/2}/2\pi\sqrt{s''_{\rm m}}.
\end{equation}
The prefactor $\nu$ can be expressed in terms of $\mathcal{Z}_2$, formally giving the
result \cite{Hanggi-00} even where the theory \cite{Hanggi-00} does
not apply.

The asymptotic technique developed in this paper allows obtaining the
prefactor $\nu$ in several limiting cases. For comparatively weak
modulation, $D\ll |s_{\rm m}| \ll R$,
Eqs.~(\ref{G_equation}), (\ref{period_averaged}) give the same result as in
Ref.~\cite{SDG}. Since the theory \cite{SDG} covers the whole range
$|s_{\rm m}| \ll R$, a transition from the Kramers limit of no
modulation to the case of arbitrarily strong modulation is now fully
described.

In the whole range where the adiabatic approximation applies,
$\Omega_F\ll 1$, we obtain
\begin{equation}
\label{C_adiabatic}
\nu=(2\pi)^{-3/2}|\mu_a\mu_b|^{1/2}D^{1/2}\omega_F(\Delta\ddot U_{\rm
m})^{-1/2}
\end{equation}
where $\mu_{a,b}$ are calculated for $t=t_{\rm m}$. Interestingly, $\nu$
(\ref{C_adiabatic}) is independent of the modulation frequency.

Eq.~(\ref{C_adiabatic}) is simplified for the modulation amplitude
close to the bifurcational value $A_c$ where the metastable and
unstable states merge. For small $\delta \! A = A_c-A$ and
$\omega_F|t-t_{\rm m}|$ the adiabatic barrier is $\Delta U(t) \propto
[\delta \!A + a_c\omega_F^2(t-t_m)^2]^{3/2}$ (here $a_c\sim A_c$), and
$|\mu_{a,b}|\propto (\delta \!A)^{1/2}$,
cf. Refs.~\onlinecite{Kurkijarvi72,Dykman80,Victora89,Tretiakov03,DGR04,Bier05}. Then,
from Eq.~(\ref{C_adiabatic}), the prefactor in the adiabatic limit
scales as $\nu\propto (\delta \!A)^{1/4}$.

The slowing down of the system motion makes the adiabatic
approximation inapplicable in the region $\delta \!A/A_c \lesssim
\Omega_F$. In contrast to the adiabatic scaling $R\propto (\delta
A)^{3/2}$, the activation energy scales here as $R\propto (\delta
\!A)^2$ \cite{DGR04}. Using the results \cite{DGR04}, we obtain from
Eq.~(\ref{G_equation})
\begin{equation}
\label{loc_nonadiabatic}
\nu = \nu_0D^{1/2}(\delta \!A)^{-1}\omega_F^{5/4},
\end{equation}
where
$\nu_0=(64\pi^7\omega_F)^{-1/4}|\partial^2_tK\,\partial^2_qK|^{1/8}/
|\partial_AK|$.  Here all derivatives are evaluated for $q,t$, and the
amplitude $A=A_c^{\rm ad}\approx A_c$ where the minimum and maximum
over $q$ of the potential $U(q,t)$ merge (once per period).

\begin{figure}[t]
\includegraphics[width=3.2in]{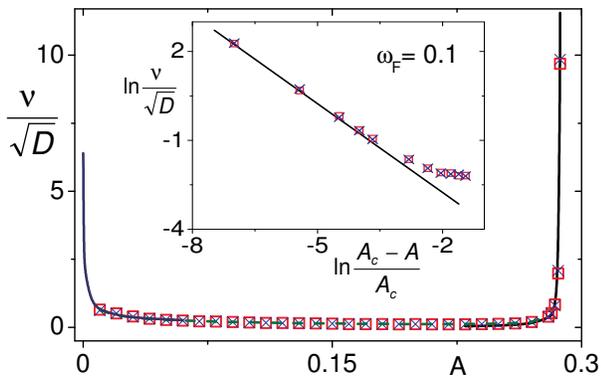}
\caption{The prefactor $\nu$ in the average escape rate $\overline{W}$
(\ref{period_averaged}). The results refer to a Brownian particle with
$K(q,t)= q^2-1/4 + A\cos(\omega_Ft)$, $\omega_F=0.1$, and describe
escape in the regime of strong synchronization, where $\nu\propto
D^{1/2}$. The solid line for small $A$ show the scaling $\nu\propto
A^{-1/2}$ \protect\cite{SDG}. The solid line for small $A_c-A$
($A_c\approx 0.29$) shows the scaling
(\protect\ref{loc_nonadiabatic}). The dashed line shows the result of
the numerical solution of Eq.~(\ref{G_equation}). The squares and
crosses show the results of Monte Carlo simulations for $R/D=5$ and
$R/D=6$, respectively.} \label{fig:prefactor_results}
\end{figure}

From Eq.~(\ref{loc_nonadiabatic}), the prefactor $\nu\propto (\delta
\!A)^{-1}$ sharply increases as the modulation amplitude approaches
$A_c$. This  is qualitatively different from the decrease of
$\nu$ in the adiabatic approximation.  The scaling $\nu\propto (\delta
\!A)^{-1}$ agrees with the numerical solution of
Eqs.~(\ref{G_equation}), (\ref{period_averaged}) for a model system
shown in Fig.~\ref{fig:prefactor_results}. The calculations in a broad
range of $A$ are also confirmed by Monte Carlo simulations.

For high frequencies, $\Omega_F\gg 1$, modulation does not lead to
exponentially strong escape synchronization. The prefactor in the
escape rate is $\nu = |\bar\mu_a\bar\mu_b|^{1/2}/2\pi$, it is
independent of the noise intensity $D$. Near the bifurcation point it
scales as in stationary systems, where $\nu \propto (\delta
\!A)^{1/2}$ and $R\propto (\delta \!A)^{3/2}$
\cite{Kurkijarvi72,Dykman80}. Very close to the bifurcation point
modulation is necessarily fast, because $|\bar\mu_{a,b}|\to 0$ for
$A\to A_c$. Therefore the prefactor always goes to zero for $A\to
A_c$. However, for small $\omega_F$ the corresponding region of
$\delta \!A$ is exponentially narrow \cite{DGR04}.

In conclusion, we have obtained a general solution of the problem of
noise-induced escape in periodically modulated overdamped systems. With
increasing modulation frequency, the pulses of escape current change from
Gaussian to strongly asymmetric; for large $\omega_F$ current modulation is
smeared out. The prefactor $\nu$ in the period-averaged escape rate is a
strongly nonmonotonic function of the modulation amplitude $A$ for low
frequencies. It first drops with increasing $A$ to $\nu\propto (D/A)^{1/2}$
\cite{SDG}, then varies with $A$ smoothly \cite{Hanggi-00,M&S-01}, and then
sharply increases, $\nu\propto D^{1/2}/(A_c-A)$ near the bifurcation amplitude
$A_c$. We found three scaling regimes near $A_c$, where $\nu\propto
(A_c-A)^{\zeta}$ with $\zeta = 1/4, -1$, or $1/2$. The widths of the
corresponding scaling ranges strongly depend on the modulation frequency.

This research was supported in part by the NSF DMR-0305746.

\end{document}